\documentclass[twocolumn,preprintnumbers,amsmath,amssymb,secnumarabic,amssymb, nobibnotes, prb]{revtex4}
\usepackage{times}
\usepackage{graphicx}% Include figure files
\usepackage{dcolumn}% Align table columns on decimal point
\usepackage{mathrsfs}
\begin{document}
\title{\Large{Ultralow-power local laser control of the dimer density in alkali-metal vapors through photodesorption}}
\author{Pankaj K. Jha$^{1,2,}$\footnote{Author to whom correspondence should be addressed. Electronic address: pkjha@physics.tamu.edu}, Konstantin E. Dorfman$^{1,3}$, Zhenhuan Yi$^{1}$, Luqi Yuan$^{1}$, Vladimir A. Sautenkov$^{1,4}$, Yuri V. Rostovtsev$^{5}$, George R. Welch$^{1}$, Aleksei M. Zheltikov$^{1,6}$, and Marlan O. Scully$^{1,2}$\footnote{Also at Baylor University, Waco, Texas 76798.}}
\affiliation{ $^{1}$Texas A\&M University, College Station, Texas 77843, USA\\
$^{2}$Princeton University, Princeton, New Jersey 08544, USA\\
$^{3}$University of California Irvine, California 92697, USA\\
$^{4}$Joint Institute of High Temperature, RAS, Moscow 125412, Russia\\
$^{5}$University of North Texas, Denton, Texas 76203, USA\\
$^{6}$M.V. Lomonosov Moscow State University, Moscow 119992, Russia}
\date{\today}
\begin{abstract}
Ultralow-power diode-laser radiation is employed to induce photodesorption of cesium from a partially transparent thin-film cesium adsorbate on a solid surface. Using resonant Raman spectroscopy, we demonstrate that this photodesorption process enables an accurate local optical control of the density of dimer molecules in alkali-metal vapors.
\end{abstract}
\maketitle
Alkali-metal vapor systems are in high demand as time and frequency standards\cite{Essen}, playing an important role in optical metrology\cite{Ramsey}, and are widely used to test fundamental principles in optical and atomic physics\cite{MOS}. Besides wide rage of applications the alkali-metal vapor is one of the most attractive and powerful model systems for laser-matter interaction, which has enabled some of the most significant discoveries in natural sciences from pioneering experimental demonstrations of radiation pressure on atoms\cite{Frisch}, optical pumping\cite{Brossel,Hawkins}, and hyperfine-structure measurements\cite{Kusch} to coherent population trapping\cite{Alzetta}, magneto-optical trapping\cite{Raab}, and Bose-Einstein condensation\cite{Anderson}. 

A routine technique for the preparation of alkali-metal vapors for a broad variety of laboratory experiments and applications is based on heated alkali-vapor cells. Alkali vapors in such cells include atomic and molecular components whose overall pressure is controlled by the temperature of the cell. Several elegant techniques have been proposed to control the densities of the atomic and molecular fractions in alkali-metal vapors. In particular, Lintz and Bouchiat\cite{Lintz} have demonstrated the laser induced destruction of cesium dimers in a cesium vapor through a quasiresonant process assisted by collisions of cesium molecules with excited-state cesium atoms and later on Ban $et\, al.$ \cite{Ban} extended this approach to rubidium. Sarkisyan $et\, al.$ \cite{Sarkisyan} also developed a simple method of thermal dissociation of cesium dimers in cesium vapor cells. 

In the past decade, laser induced atomic desorption (LIAD)\cite{Karaulanov, Alexandrov, Graf} technique has gain much attention for controlling the atomic density in cells coated with paraffin etc., In such cells the atoms get adsorbed on the surface of the vapor cell.  In a typical LIAD experiment, a desorption laser illuminates a coated vapor cell and its effect is studied by the analyzing the absorption/transmission of a weak probe field resonant to some atomic transition of the alkali vapor. Work related to this area has been primarily focused on controlling the atomic densities for e.g. Rb, Cs, K, Na etc. First initiative in the direction of control over dimer concentration using LIAD was studied by the Berkeley group\cite{Acosta}. 

\begin{figure}[b]
\centerline{\includegraphics[height=6.0cm,width=0.44\textwidth,angle=0]{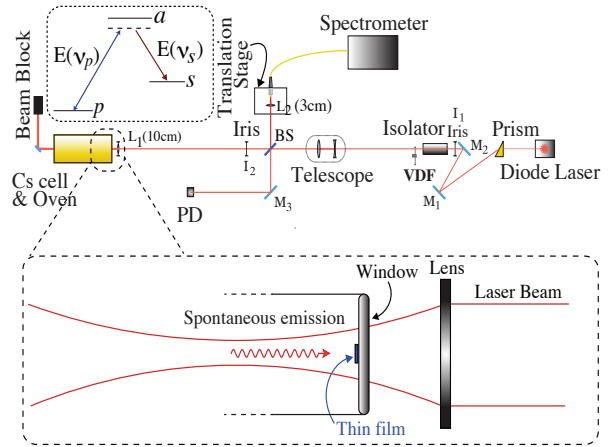}}  
 \caption{Experimental setup. The upper inset shows a simple three-level model for Raman scattering. Here the lower two levels $p$ and $s$ and upper level $a$ are the vibrational states the ground state X$^{1}\Sigma^{+}_{\text{g}}$ and excited state B$^{1}\Pi_{\text{u}}$ of cesium dimer respectively. The lower inset shows the zoomed part near the window. A thin film of metallic cesium is condensed on one side of the cell inside the oven. The Raman signal generated in the backward direction is collected and analyzed using the spectrometer. VDF  is variable density filter; L is lens and BS is beam splitter.}
 \end{figure}  
In this Letter, we extend the laser-induced photodesorption technique to ultralow laser power and use resonant Raman spectroscopy to demonstrate that LIAD\cite{Gozzini, Mariotti} enables an accurate local control of the dimer density in alkali-metal vapors. Our experimental strategy is based on studying the backscattered Raman signal from the alkali-metal vapor while illuminating a thin film of metal deposited on the window of an uncoated vapor cell  using continuous wave laser at milli-watt power [see inset Fig. 1]. In our experiment we used a cylindrical uncoated Pyrex cell with a diameter of 3 cm and a length of 7.5 cm. After desorption from the film the alkali monomers (atoms) can form dimers, trimers and higher order oligomers by colliding with each other. Possibility of dimers adsorption on the surface of the film is beyond the scope of the current work.

Our experimental setup is shown in Fig.1. A tunable free-running single-mode diode laser [Sanyo DL7140-201] was used for spectroscopy of cesium dimers.  While the laser wavelength of the diode laser can be set coarsely by adjusting the temperature (+0.04nm/K), fine frequency tuning was performed by varying the injection current (-0.04cm$^{-1}$/mA). The input laser beam was collimated by an aspheric lens, and a prism was used to compress the beam size along the horizontal axis to make it circular. Unfocussed and collimated beam diameter was $\sim 3$mm. The telescope system was further introduced to expand the beam diameter by a factor of 2 and a variable density filter (VDF) controlled the intensity of the beam. The collimated laser beam was focused into the cell using a lens L$_{1}$ [ focal length $f$=10 cm] into the cell. A circular thin film of cesium was deposited on the inner surface of the cell window at a distance of $\sim3$cm from the lens. The backscattered Raman signal was collected into a multimode fiber which conducts the light into a diffraction spectrometer [Ocean Optics HR2000: spectral resolution 0.065 nm]. Irises were used to collimate the beams and block diffused scattered radiation due to reflections from the cell windows and other optics. 

 \begin{figure}[t]
\centerline{\includegraphics[height=8.0cm,width=0.40\textwidth,angle=0]{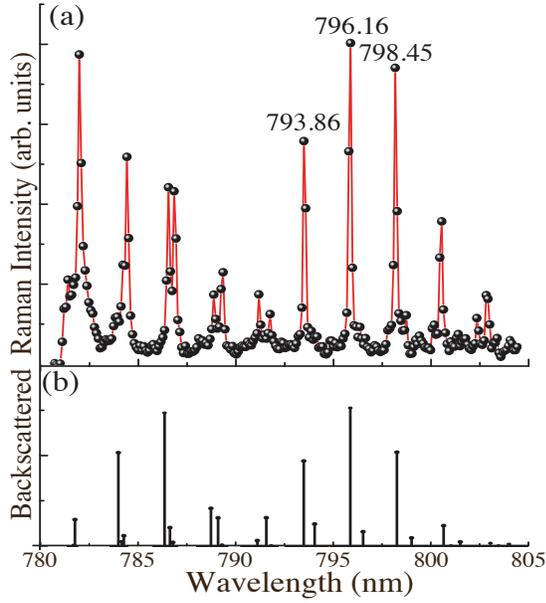}}  
 \caption{Raman spectra in the backward direction (in arbitrary units) (a) experimental and (b) numerical simulation (discussed in the text).}
 \end{figure}
The laser wavelength was set resonant to the electronic transition  X$^{1}\Sigma^{+}_{\text{g}} \leftrightarrow \text{B}^{1}\Pi_{\text{u}}$ of the cesium dimer. The absorption band of the transition X$^{1}\Sigma^{+}_{\text{g}} \leftrightarrow \text{B}^{1}\Pi_{\text{u}}$ ranges from 755nm to 810nm\cite{Gupta78}.  In Fig. 2(a), we have shown one such spectrum collected in the backward direction. We tuned the pump laser wavelength to the resonance\cite{DB} by observing the intensity of one of the Raman peaks (796.16nm). The maximum value of the intensity corresponds to pump wavelength $\lambda_{p}=779.9010$ nm (air) [WA-1500 wave meter from Burleigh].  In Fig. 3. we have plotted the resonance enhancement of the peak ($796.16$nm) against the one photon detuning $\Delta= \omega_{ap}-\nu_{p}$ which indicates the high sensitivity of the Raman response to the pump wavelength. To simulate the spontaneous Raman spectral response we used~\cite{mukamel}
\begin{equation} \label{Raman}
\begin{split}
&S_{RAMAN} (\nu_p,\nu_s) \\
&= 2 \pi \sum\limits_{p,s} P(p)|\chi_{sp}(\nu_p)|^2 \delta(\omega_{sp}+\nu_s-\nu_p),
\end{split}
\end{equation}
where
\begin{eqnarray}
\chi_{sp}(\nu_p) = \sum\limits_{a} \frac{\wp_{sa}\wp_{ap}}{-\omega_{ap}+\nu_p+i\Gamma}.
\end{eqnarray}
\begin{figure}[t]
\centerline{\includegraphics[height=5.4cm,width=0.43\textwidth,angle=0]{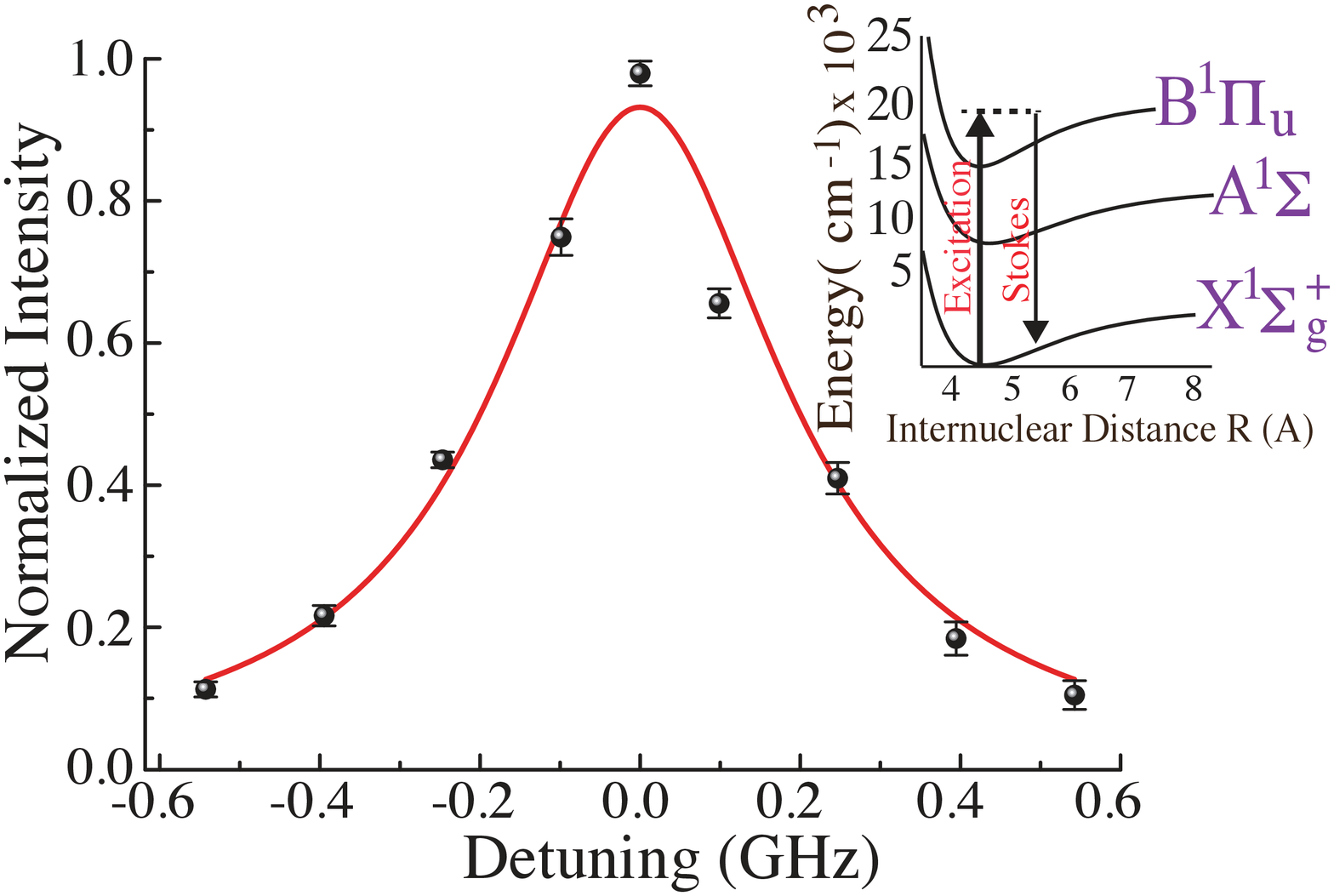}}  
\caption{Resonance enhancement of the Raman peak at 796.16nm as a function of the one-photon detuning. Full width at half maximum value ($\sim 0.3$ GHz) is consistent with the Doppler broadening from the vapor phase. Insert shows the energy levels of Cesium dimer relevant to our experiment.}
\label{scheme}
\end{figure}
Here $\nu_{p}$ and $\nu_{s}$ are the pump and the Stokes frequency  respectively. $P(p)$ is the normalized thermal population distribution given as $P(p) = e^{-E_p/kT}/\sum\limits_{p}e^{-E_{p}/k_BT}.$ $\hbar \omega_{ij}$  and $\wp_{ij}$ are the energy difference and the transition dipole moment between the levels $i$ and $j$ respectively. We have approximately calculated Franck-Condon factors (FCF) by using the exact eigenfunctions of the Morse Potential~\cite{morse}. $\Gamma$ is the transverse relaxation rate. $E_v =\hbar \omega (v + \frac{1}{2}) - h \omega\chi (v +\frac{1}{2})^2$ is the energy of vibrational level $v$, where $\omega$ is the vibrational frequency and $\omega\chi$ is the vibrational anharmonicity~\cite{steele}. For cesium\cite{GH} ground state X$^{1}\Sigma^{+}_{\text{g}}$, $\omega_{g} \sim 42.20$$\left(\text{cm}^{-1}\right)$ and $\omega_{g}\chi_{g} \sim 0.0819$$\left(\text{cm}^{-1}\right)$ while in the excited state B$^{1}\Pi_{\text{u}}$, $\omega_{e} \sim 34.33$$\left(\text{cm}^{-1}\right)$ and $\omega_{e}\chi_{e} \sim 0.08$$\left(\text{cm}^{-1}\right)$. Different amplitudes of the FCFs for different transitions between the vibrational levels of the electronic states X$^{1}\Sigma^{+}_{\text{g}}$ and B$^{1}\Pi_{\text{u}}$ indicate that the dipole moment for different transition has different magnitude\cite{Luqi} as the square of the dipole moment is proportional to the FCFs. Consequently the gain for different transitions is different. Fig. 2(b) shows the simulated spectrum in the Stokes region using Eq.(\ref{Raman}) which is an excellent agreement with the experimental data shown in Fig. 2(a).

The main result of our work is shown in Fig. 4 where we have plotted the intensity of Raman peak (796.16nm) as a function of the pump power for different cell temperatures. Here curves 1, 2 and 3 correspond to the cell temperature $T_{c}$= $513$K, $526$K and $543$K respectively. In our experiment we also monitored the transmission through the metal film before and after the measurements of laser induced fluorescence (LIF) from cesium vapor. The linear dependence between transmitted power and input power as shown in Fig. 4 (inset) indicates that under our experimental conditions the transmission through the film is independent of the pump power. As the fluorescence signal depends on the input power which indicates that the laser light induce desorption of cesium atoms from the metal film. Thus, power independence of the film transmission can be explained by moderate reduction of the film, of the order of  several monolayers. The efficiency of the desorption increases with the cell temperature.  

\begin{figure}[t]
\begin{center}
\centerline{\includegraphics[height=7.0cm,width=0.41\textwidth,angle=0]{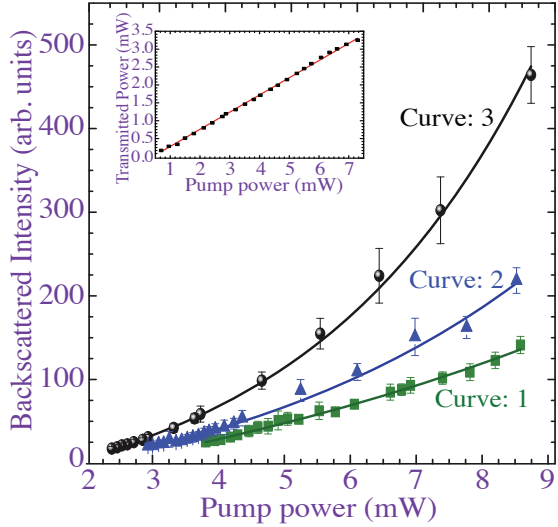}}  
\end{center}
\caption{Plot of the backscattered intensity (arb. units) of the Raman peak at 796.16nm vs the pump power for three different choices of the cell temperature in the presence of the film. Dots illustrate the experimental data and solid lines are fitting using Eq.(\ref{Fit}).}
\label{Main}
\end{figure} 
To fit our experimental data, we assumed the following fitting function
\begin{equation}\label{Fit}
I=\sum_{n=1}\alpha_{n}P^{n}
\end{equation}
the coefficients $\alpha_{n} (n=1,2,3...)$ contains the information about the number density of the dimers, differential cross-section, geometry of the gain medium, contribution due to photodesoprtion etc. In the absence of the film $\alpha_{n}=0$ for $n \geq 2$. We further normalize Eq.(\ref{Fit}) with respect to the linear contribution $(I_{1}=\alpha_{1}P)$, which yields
\begin{equation}\label{NFit}
\frac{I}{I_{1}}=1 +\beta_{1} P +\beta_{2}P^{2} +...
\end{equation}
where $\beta_{n}=\alpha_{n+1}/\alpha_{1}$. Next we simplify our analysis by considering $n=1$ term only. To account for the background noise we added $I_{0}$ in Eq.(\ref{Fit}). Generally, the intensity of the Stokes radiation from a volume of the medium of unit area and a length $dz$ is given by~\cite{Bloem}
\begin{equation}
dI=N_{0}(T_{c})\frac{d\sigma}{d\Omega}\zeta P dz
\end{equation}
where $N_{0}(T_{c})$ is the density of the scattering molecules, $d\sigma/d\Omega$ is the differential cross section of the spontaneous Raman scattering, $\zeta$ is the solid angle in which the scattering is observed, and $P$ is the power of the laser radiation. In table I we present the fitting parameter $\beta_{1}=\alpha_{2}/\alpha_{1}$ and estimated the number density of the cesium dimers. The numbers in the parentheses corresponds to the error in the fitting parameters. Here,  $T_{c}$ is the cell temperature and $N_{1}$ is the number density of the dimers when the pump power is $P \sim 8.5$mW.  In order to estimate for $N_{1}$ we use
\begin{equation}
N=N_{0}\left(1+\beta_{1}P\right)
\end{equation}
From the estimated values at $T_{c} =543$K and $P \sim 8.5$mW, the number density of the cesium dimers is $\sim 6$ times larger than in the absence of atomic desorption from the film. We observed this enhancement in dimer density even at lower cell temperature $T_{c}=513$K. Let us introduce an effective temperature $T_{e}$ which is equivalent to the cell temperature at  which the number density of cesium dimers is $N_{e}=N_{1}$. Using the vapor pressure formula \cite{Nesmeyanov}, we estimated for $T_{e}$ and the results are shown in Table I. We see that the effective temperature can be as high as $\sim 54$K above the cell temperature.

\begin{table}[t]
\caption{Numerical values of the fitting parameter $\beta=\alpha_{2}/ \alpha_{1}$ and the number density of the Cs$_{2}$ dimers at maximum pump power $P\sim 8.5$mW.}
\vspace{1mm}
\centering
\begin{tabular}{c c c c c}
\hline\hline
Curve  & $\text{T}_{c}$ \text{(K)} & $\beta_{1}$& $\text{N}_{1}/\text{N}_{0}$ &$\text{T}_{e}$ (K)\\ [0.5ex]
\hline
1 &513&0.1734(0.009)&2.474(0.108)&567 \\
2 &526&0.2972(0.016)&3.526(0.421)& 578 \\
3 &543&0.6704(0.025)&6.698(0.267)& 597 \\[1ex]
\hline\hline
\end{tabular}
\label{Table1}
\end{table}
For all the estimates, we have assumed that the spontaneous Raman is the dominant phenomena here. To verify the assumption that the nonlinear behavior of the Raman signal is not due to stimulated Raman scattering(SRS), we estimate the gain coefficient for SRS under our experimental conditions. For SRS the Stokes intensity in the backward direction, assuming that pump intensity is not depleted, is given by\cite{Bloem} 
\begin{equation}
 \frac{d}{dz}I^{b}_{s}(z)=-gI^{b}_{s}(z)I_{p},
\end{equation}
where the gain coefficient\cite{Bloem, Thesis} 
\begin{equation}
g =\left(\frac{N |\wp_{ap}|^{2}|\wp_{as}|^{2}\nu_{s}n^{(0)}_{ps}}{2\epsilon^{2}_{0}c^{2}\hbar^{3}\Delta^{2}\Gamma}\right).
\end{equation}
Here $n^{(0)}_{ps}=\varrho^{(0)}_{pp}-\varrho^{(0)}_{ss}$ is the population inversion and $\Gamma$ is the dephasing of the Raman coherence. In the temperature range from 470 K - 540 K the atomic $N_{a}$ and the molecular number density $N_{m}$ lies in the range $10^{15}-10^{16}$ cm$^{-3}$ and $10^{13}-10^{14}$ cm$^{-3}$ respectively. The ratio\cite{Nesmeyanov} of the atomic density $N_{a}$ to the molecular density $N_{m}$ is order of $10^{2}$. In the experiment we optically pumped the cesium dimers at $T \sim 545$K and density $2\times 10^{14}$cm$^{-3}$. The wavelength of the laser was tuned to $\lambda_{p} =779.90$nm. The diameter of the focused beam at the waist is $d = 4\lambda_{s}f/\pi D \sim 34 \mu$m, where the unfocused beam diameter $D$ and the focal length of the lens $f $ are 0.6cm and 10cm respectively. The depth of the focus is $L=8\lambda_{p}f^{2}/\pi D^{2} \sim 0.11$cm. The pump intensity at the waist is $I_{p} \sim 300$W/cm$^{2}$. The differential spontaneous cross section was estimated as $d\sigma/d\Omega \sim 3 \times 10^{-21}$cm$^{-2}$. For the resonance enhanced Raman, the Doppler broadening $\Delta_{D} =k_{p}v_{th}\sim 2\times 10^{9}$s$^{-1}$ and we assumed the decay rate of Raman coherence $\Gamma \sim1$GHz. From Eq.~(7) and the experimental parameters we obtained $g \sim 1.2\times 10^{-2}$W$^{-1}$cm. Hence the estimate for $gI_{p}L\sim 0.4 $ clearly indicates that the contribution from SRS can be safely neglected.

To summarize, we used ultra-low power continuous-wave (cw) diode laser to optically control the density of cesium dimers through photodesorption from a thin film of cesium in an uncoated Pyrex cell. To probe the dimer concentration, we collected the spontaneous Raman signal in the backward direction which serves the two-fold purpose (a) the signal is from the dimers and (b) envision the idea of remote detection of chemicals using ultra-low power cw lasers. We observed a nonlinear behavior [as shown in Fig.~\ref{Main}] in the peak intensity against the pump power contrary to the linear dependence behavior well known from the spontaneous Raman theory.  The deviation from the linear behavior is due to the contribution of the Raman signal generated from the cesium dimers produced through photodesorption from the thin film on the window. Under the experimental conditions we estimated the number density of the dimers increased substantially in the presence of the film. 

The main goal of this work is to make a significant step in the direction of LIAD technique, which offers a powerful control over atomic/dimer density in coated cells. An optical control over the vapor density, as shown here, will offer an additional tool for numerous applications of the alkali-metal vapors for eg. time and frequency standards\cite{Essen}, optical metrology\cite{Ramsey}, test fundamental principles in optical and atomic physics\cite{MOS}, as well as to be the most attractive and powerful model systems of laser-matter interaction.

It is our pleasure to thank David Lee and Dudley Herschbach for insightful discussions. We also gratefully acknowledge support for this work by National Science Foundation Grant EEC-0540832 (MIRTHE ERC), the Office of Naval Research, the Robert A. Welch Foundation (A-1261), the Seventh European Framework Programme (CROSS TRAP 244068 project), the Russian Foundation for Basic Research. P. K. Jha,  Z. Yi and L. Yuan are supported by the Herman F. Heep and Minnie Belle Heep Texas A\&M University Endowed Fund held/administered by the Texas A\&M Foundation and Welch Foundation Graduate Fellowship.


\begin{thebibliography}{99}
\bibitem{DB} The width of the response of Raman signal against the single photon detuning will be governed by Doppler broadening.
\end{thebibliography}
\end{document}